\documentclass[a4paper,conference]{IEEEtran}
\usepackage{algorithm}
\usepackage{color}
\usepackage{algpseudocode}
\usepackage{graphics}
\usepackage{epsfig}
\usepackage{cite}
\usepackage{threeparttable}
\usepackage{graphicx}
\usepackage{picinpar}
\usepackage{amssymb}
\usepackage[cmex10]{amsmath}
\usepackage{subfigure}
\usepackage{subfig}
\usepackage{stfloats}

\usepackage{bm}
\usepackage{url}
\usepackage{booktabs}
\usepackage{setspace}
\input epsf

\begin{document}
	
\title{\huge{Reinforcement-Learning-Enabled Beam Alignment for Water-Air Direct Optical Wireless Communications}}

%Optical Beam Alignment based on Reinforcement learning in Direct Water-Air Communication System
\author{
Jiayue~Liu$^{1,2,3}$, Tianqi~Mao$^{1,2,3}$, Dongxuan He$^{4}$, Yang Yang$^{5}$, Zhen Gao$^{1,2,3}$, Dezhi Zheng$^{1,2,3}$, Jun Zhang$^{1,2}$  

 \\$^1$State Key Laboratory of CNS/ATM, Beijing Institute of Technology, Beijing 100081, China
 \\$^2$ MIIT Key Laboratory of Complex-Field Intelligent Sensing, Beijing Institute of Technology, \\Beijing 100081, China \\$^3$ Yangtze Delta Region Academy, Beijing Institute of Technology (Jiaxing), Jiaxing 314019, China \\$^4$ School of Information and Electronics, Beijing Institute of Technology, Beijing 100081, China \\$^5$ Beijing Key Laboratory of Network System Architecture and Convergence, School of Information and\\ Communication Engineering, Beijing University of Posts and Telecommunications, Beijing 100876, China\\
E-mails: $\{$jiayue\_liu@bit.edu.cn, maotq@bit.edu.cn, dongxuan\_he@bit.edu.cn, \\yangyang01@bupt.edu.cn, gaozhen16@bit.edu.cn, zhengdezhi@bit.edu.cn, buaazhangjun@vip.sina.com$\}$ 
\vspace{-4mm}}

\maketitle
\begin{abstract}
The escalating interests on underwater exploration/
reconnaissance applications have motivated high-rate data transmission from underwater to airborne relaying platforms, especially under high-sea scenarios. Thanks to its broad bandwidth and superior confidentiality, Optical wireless communication has become one promising candidate for water-air transmission. However, the optical signals inevitably suffer from deviations when crossing the highly-dynamic water-air interfaces in the absence of relaying ships/buoys. To address the issue, this article proposes one novel beam alignment strategy based on deep reinforcement learning (DRL) for water-air direct optical wireless communications. Specifically, the dynamic water-air interface is mathematically modeled using sea-wave spectrum analysis, followed by characterization of the propagation channel with ray-tracing techniques. Then the deep deterministic policy gradient (DDPG) scheme is introduced for DRL-based transceiving beam alignment. A logarithm-exponential (LE) nonlinear reward function with respect to the received signal strength is designed for high-resolution rewarding between different actions. Simulation results validate the superiority of the proposed DRL-based beam alignment scheme.

% abstract 
%abstract部分概要
%当前水-空跨域通信发展，跨域直连激光通信技术亟待研究。
%本文提出的方法
%采用海浪谱（wave spectrum）方法对海浪进行建模，进而建立光路（光信道）模型。
%在光路模型的基础上用深度强化学习（DDPG算法）进行训练，将光路模型作为环境（environment），将接收光强作为奖励（reward），以实现波束对准，获得尽可能大的接收光强。
\end{abstract}


\vspace{-3mm}
%\addtolength{\topmargin}{0.1cm}

\section{Introduction}\label{s1}

% Introduction 部分
% 1 水下光通信、跨域光通信的研究的背景与系统特性描述
% 2 目前关于水下光通信、跨域光通信的研究成果
% 3 本篇文章的研究内容简要介绍

% 1.水下与跨域通信背景与优势
% 水下和空中平台的通信很重要（说明为什么很重要）

Recent years have witnessed unprecedented developments of maritime technologies including unmanned underwater vehicles (UUVs) and underwater buoy platforms \cite{An_overview}. Such advancements have facilitated numerous civilian/military applications such as ocean exploration and tactical surveillance \cite{An_overview,Overview_UWC}. To guarantee timely backhauling of the measurement data, establishing communication links between underwater and airborne relaying platforms, i.e., water-air links, can be mandatory, especially for high-sea scenarios, where the longshore stations are too distant to support direct transmission. Traditional approaches tend to employ acoustic communications attributed to its robustness to the deleterious propagation channel in underwater environment, which is constrained by limited bandwidth and excessive latency \cite{Zhou_access_19}. On the other hand, despite the broad achievable communication bandwidth under terrestrial scenarios, the radio-frequency (RF) signal suffers from severe attenuation/absorption during underwater propagation, making it inapplicable for practical implementations \cite{A_Survay_UOWC}. Alternatively, optical wireless communication (OWC) has been demonstrated to support superior throughput levels for underwater and free-space transmission, thanks to its substantial unlicensed spectrum resources and moderate propagation loss for both water and atmospheric mediums \cite{Waving_Effect}. Therefore, OWC has become one promising candidate for next-generation broadband water-air data transmission.

%annnoted by Mao%With the development of marine science and engineering technology, various under water equipment such as Unmanned Underwater Vehicle (UUVs) and under water buoys are deployed to the distant sea. These devices perform various tasks, such as ocean exploration and reconnaissance. In order to transmit the exploration result in time, high rate data transmission method is required for under water devices to communicate with airborne device which is accessible to free space transmission links. In this case, water-air communications has become an important research topic. 
% 由于水下环境的特殊性，传统的微波通信效果不好 -> 传统方法多使用声波 -> 声波的问题（比如带宽）
%annoted by Mao% Due to the special under water environment, radio frequency communication lacks good performances, so that traditional method always use acoustic communication.\cite{Underwater_and} However, acoustic communication is severely limited by bandwidth and cannot transmit information efficiently. 
% 与这两种方法相比，激光有xxx优势，受到学术和产业界的广泛关注
%annnoted by Mao%Compare to the two methods upon, optical wireless communication (OWC) has a better performance, for it transmits much further than radio wave, and has much wider bandwidth than acoustic wave as well.\cite{Underwater_and} Additionally, optical communication has much better performance in cross-domain communication than radio and acoustic methods. Therefore, OWC has received widespread attention from academia and industry.

% 2.现有跨域通信对比
The majority of existing literature on water-air OWC concerns multi-hop communications with offshore relaying platforms \cite{Recent_Progress,xiaoyang2019performance,Underwater_optic,Effect_of}, which cleverly circumvents penetration through highly dynamic water-air interface. However, such methods become inapplicable without the presence of relaying platforms. This can happen for high maneuverability tasks where deployment of relaying platforms cannot be instantly accomplished, and for denied environments where offshore relaying nodes have been destroyed. Therefore, it is worthwhile to investigate direct OWC across the water-air interface as a complement to relaying strategies.

Unlike the relaying counterpart, the direct water-air optical transmission suffers from severe transceiving beam alignment caused by the dynamic characteristics of the water-air interface. To be more specific, refraction of the wavy water surface can cause random attenuation and deflection to the optical path, leading to frequent outages, especially for highly directional laser transmission \cite{Underwater_and}. 
% 引用一些文章，说明实验情况下的直连通信
There have been preliminary researches working on water-air direct OWC \cite{Mitigation}. applied photodiode array to detect the beam-direction changes caused by waves, and used micro-electro-mechanical system (MEMS) to compensate for the beam misalignment. However, this method makes it difficult to deal with horizontal offset when the transmission distance is sufficiently large.
Besides, \cite{Preliminary} utilized the scattering of the underwater LED emitter to ensure reliable water-air transmission, which was verified by experimental demonstrations.
Afterward, the authors further investigated the waving effect on channel gain of water-air OWC, and introduced array-based transceivers to enhance the achievable rate and the error performance \cite{Waving_Effect}. Note that the proposed methods only mitigated the impacts of dynamic waves passively, which may cause unstable channel gain and even random interruptions without active beam alignment operations.
%Moreover, this method is also lack of ability to handle non-aligned environment, for the achievable rate decreases sharply in these situations.

% 为此，提出了xxx技术 -> 收尾
From the aforementioned discussions, the previous breakthroughs mainly concentrate on hardware implementations of water-air direct OWC. On the other hand, there is lack of research on effective beam alignment strategies for water-air OWC.
To fill the gap in related research, this article proposes a beam alignment algorithm for water-air direct OWC based on deep reinforcement learning (DRL). Specifically, the water-air OWC channel is mathematically modeled based on the wave spectrum theory, followed by characterization of the optical channel using ray-tracing method. 
Meanwhile, a DRL environment with a designed reward function is established on basis of the proposed channel model, in which the beam alignment algorithm is trained utilizing deep deterministic policy gradient (DDPG) strategy.
Simulation results demonstrate that the proposed beam alignment method has superior performance in keeping high channel gain and resisting channel variations.
%{\color{red}(please refer to abstract and conclusion)}

\section{System Model}\label{s2}

% 第二部分，阐明信道仿真模型
% 1 跨域信道仿真环境的原理
% 2 海浪仿真以及影响
% 3 信道整体特征
% 4 信道模型的建立与仿真

% system model
In this section, the mathematical model of the propagation channel of water-air direct OWC system is provided.

\subsection{Water-Air Communication Scenario}\label{s2.1}
\begin{figure}[!t]
    \centering
    \includegraphics[width=0.85\linewidth, keepaspectratio]{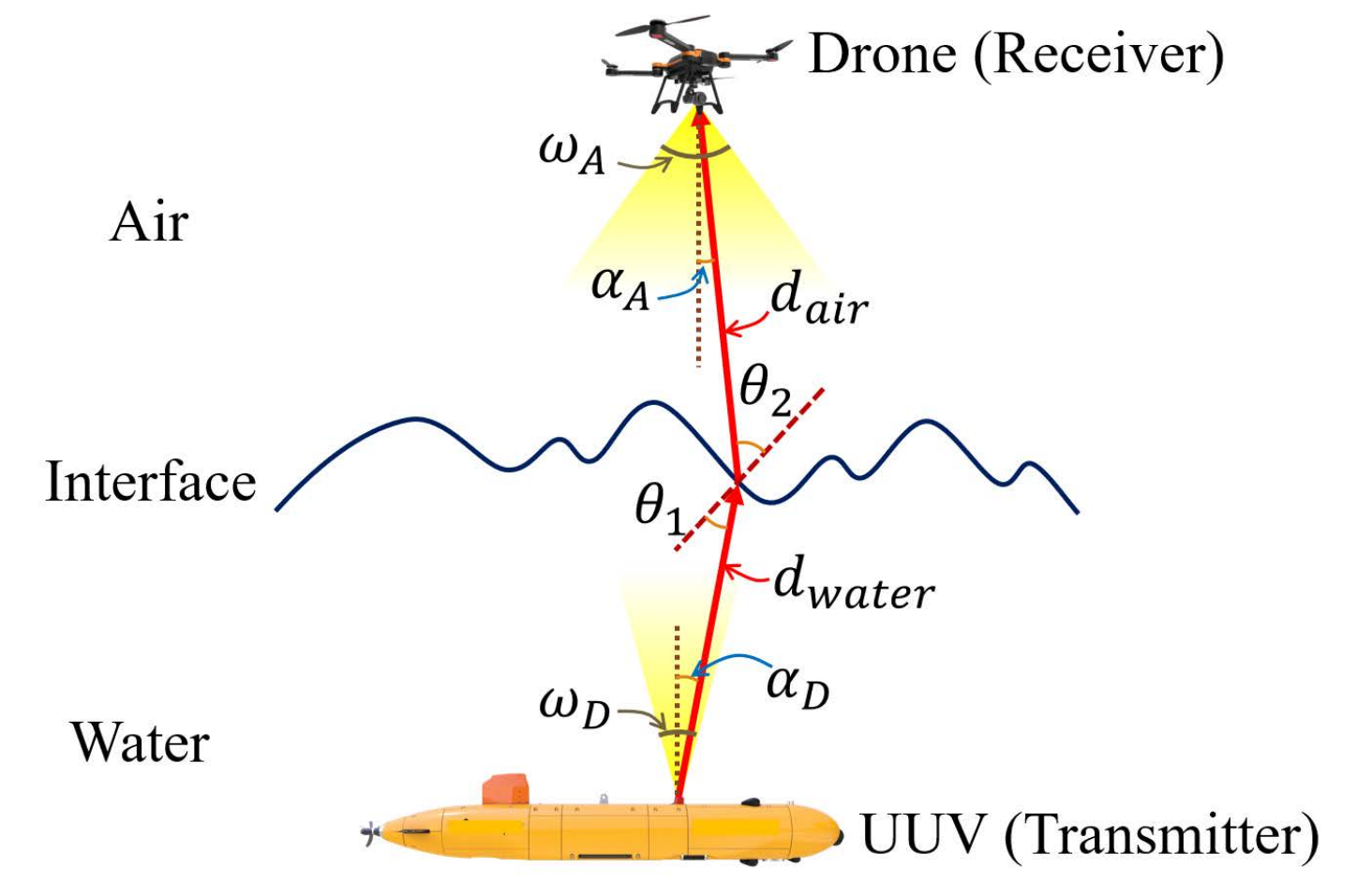}
    \caption{Illustration of water-air direct optical wireless communications between underwater and airborne platforms.}
    \label{fig:Environment_model}
    \vspace{-6mm}
\end{figure}

% 场景:图，水下潜器->无人机，海浪，定义变量，简化
As illustrated in Fig. \ref{fig:Environment_model}, this article considers uplink direct OWC between UUVs and airborne drones, i.e., water-air OWC for brevity. Under this scenario, the optical signals are transmitted from the laser diodes (LD) through the water and atmosphere mediums sequentially, and detected by the avalanche photodiode (APD) at the airborne receiving platform. As presented in Fig. \ref{fig:Environment_model}, to characterize the propagation channel of water-air OWC, we define the maximum accessible angle, the maximum angle that the transmitter/receiver can emit/detect optical beam,
%\footnote{\color{red}give definitions of maximum accessible angle and cite a reference}
of LD and APD as $\omega_D$ and $\omega_A$, respectively \cite{Improvement_of, Study_on}. Besides, the angles of departure and arrival are denoted as $\alpha_D$ and $\alpha_A$, and the propagation distances through water and atmosphere mediums are represented by $d_{water}$ and $d_{air}$. Moreover, the optical signals are assumed to cross the water-air interface at the incident angle $\theta_1$ and emergence angle $\theta_2$. 

\subsection{Channel Model}\label{s2.3}
The propagation channel of water-air OWC is determined by characteristics of the LD and APD, the path loss through water and air mediums, and the penetration loss crossing the water-air interface. Hence, the optical channel gain can be formulated as
\begin{equation}\label{func5}
    G = G_D(\alpha_D)\cdot G_p(d_{water},d_{air})G_{ref}(n,\theta_1)\cdot G_A(\alpha_A)
\end{equation}
where $G_D$, $G_A$ denote the departure and arrival gains, and $G_p$ and $G_r$ stand for the path gain and refraction gain, respectively. 

\subsubsection{Departure Gain}
$G_D$ depends on the departure angle $\alpha_D$ and the LD wavelength $\lambda$, calculated as \cite{Improvement_of}
% {\color{red}\cite{Improvement_of}}
\begin{equation}\label{func6}
    G_D(\alpha_D) = \exp\left(-\frac{2\sin^2(\alpha_D)}{\omega_D^2[1+(\frac{\lambda  \cos(\alpha_D)}{\pi\omega_D^2})^2]}\right)
\end{equation}

\subsubsection{Path Gain}\label{s2.3.2}
The value of $G_p$ is decided by the spreading loss and absorption effects through water and air mediums, which is expressed as \cite{Underwater_and}
\begin{equation}\label{func7}
    G_p(d_{water},d_{air}) = \frac{\exp(\alpha(\lambda)d_{water})}{(d_{water}+d_{air})^2}
\end{equation}
where $\alpha(\lambda)$ denotes the absorption coefficient according to Lambert's law \cite{Absorption_spec}. For simplicity, we omit bubbles or turbulence in the water medium.

\subsubsection{Refraction Gain}\label{s2.3.3}
According to Snell's Law and Fresnel Equation, $G_{ref}$ can be calculated with the incident angle $\theta_1$ and the refraction indices $n_1$ and $n_2$ for water and air mediums, shown as \cite{Fresnel}
\begin{multline}\label{func8}
    G_{ref} = 1-\frac{1}{2}\bigg[(\frac{n_2\cos\theta_1-n_1\cos\theta_2}{n_2\cos\theta_1+n_1\cos\theta_2})^2 
    \\+ (\frac{n_1\cos\theta_1-n_2\cos\theta_2}{n_1\cos\theta_1+n_2\cos\theta_2})^2\bigg]
\end{multline}

\subsubsection{Arrival Gain}\label{s2.3.4}
The value of $G_A$ can be determined by the maximum accessible angle of the APD $\omega_A$ and the arrival angle $\alpha_A$, written as \cite{Study_on}
\begin{equation}\label{func9}
    G_A = \frac{n_2^2}{\sin^2\omega_A}\cos\alpha_A
\end{equation}

%In conclusion, the complete channel function can be obtained by multiplying the above terms.
% 总方程（这玩意还要不要加呢？）
%\begin{multline}\label{func10}
  %  G = \exp(-\frac{2^2\cos^2(\alpha_D)}{\omega_D^2[1+(\frac{\lambda \cos(\alpha_D)}{\pi\omega_D^2})^2]}) \cdot \frac{\exp(\alpha(\lambda)d_{water})}{d_{air}^2}\\ \cdot (1-\frac{1}{2}\bigg[(\frac{n_2\cos\theta_1-n_1\cos\theta_2}{n_2\cos\theta_1+n_1\cos\theta_2})^2 + (\frac{n_1\cos\theta_1-n_2\cos\theta_2}{n_1\cos\theta_1+n_2\cos\theta_2})^2\bigg])\\ \cdot \frac{n^2}{\sin^2\omega_A}\cos\alpha_A
%\end{multline}
%The channel gain shown as function \ref{func10} is the most important factor to judge the performance of the OWC systems.
% $$H = \frac{A(m+1)}{2\pi{d_{air}}^2{d_{water}}^2} \cos^m(\alpha_D)G(\alpha_A)\cos(\alpha_A)G_{ref}(n,\theta_1)$$

\section{Mathematical Modelling of Optical Paths Crossing Water-Air Interface}\label{s3} 
Different from free-space/underwater optical communications, the water-air OWC channel would be significantly impacted by dynamic refraction effects of the waving water surface. Therefore, accurate mathematical modeling of the impacts of the dynamic water surface on optical path can be necessary for channel characterization of water-air OWC. Without loss of generality, this paper mainly investigates the dynamic characteristics of ocean waves. Below the mathematical model of the waving water-air interface is derived with the inspiration of existing oceanography theories. Then the optical propagation path can be determined using ray-tracing based on the established model.

\begin{figure*}[!t]
	\centering
	\subfigure[Illustrations of the JONSWAP Spectrum.]{
		\begin{minipage}[t]{0.4\linewidth}
			\centering
			\includegraphics[width=1\linewidth]{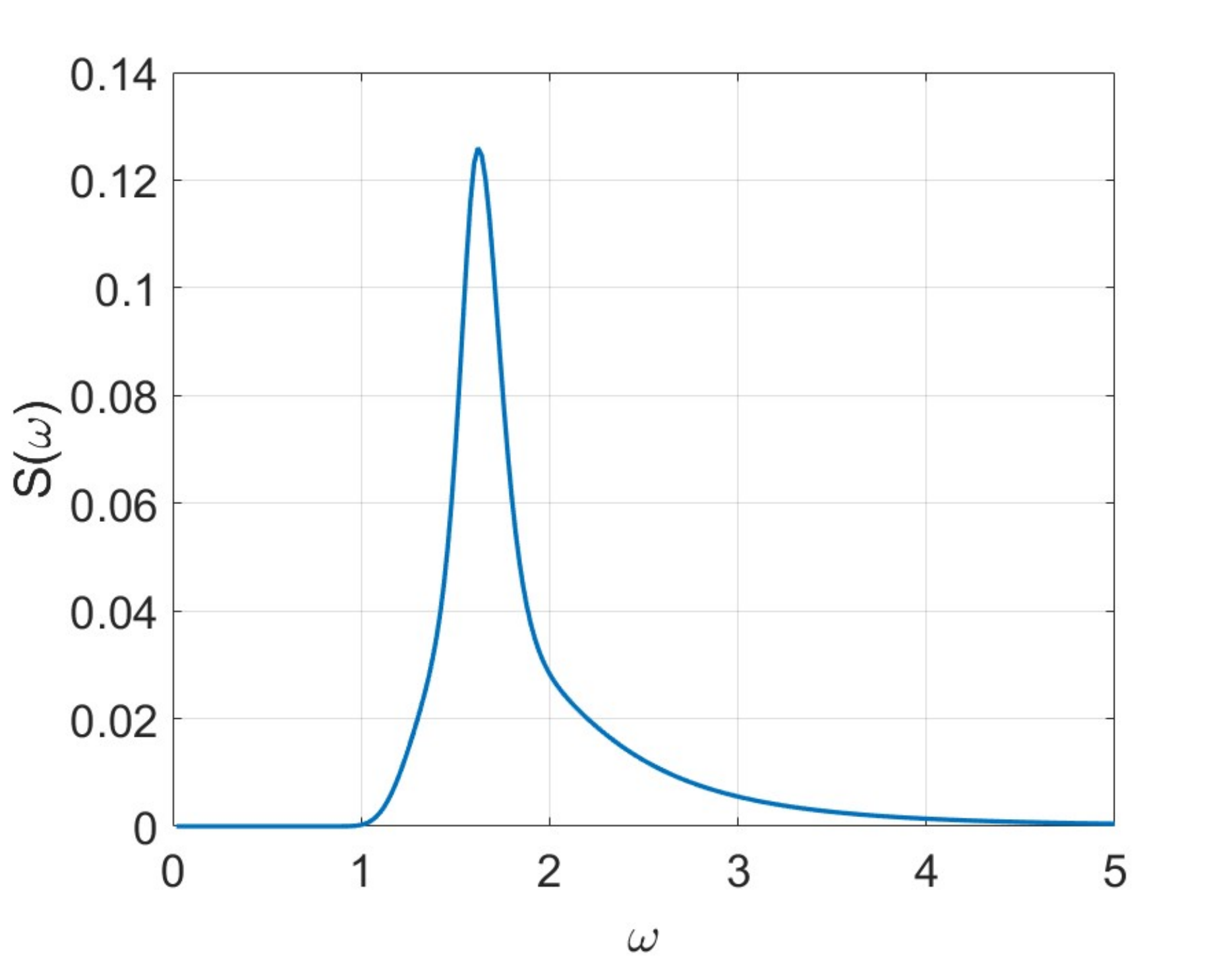}
		\end{minipage}
	}%
	\subfigure[3-D illustrations of the JONSWAP Spectrum.]{
		\begin{minipage}[t]{0.4\linewidth}
			\centering
			\includegraphics[width=1\linewidth]{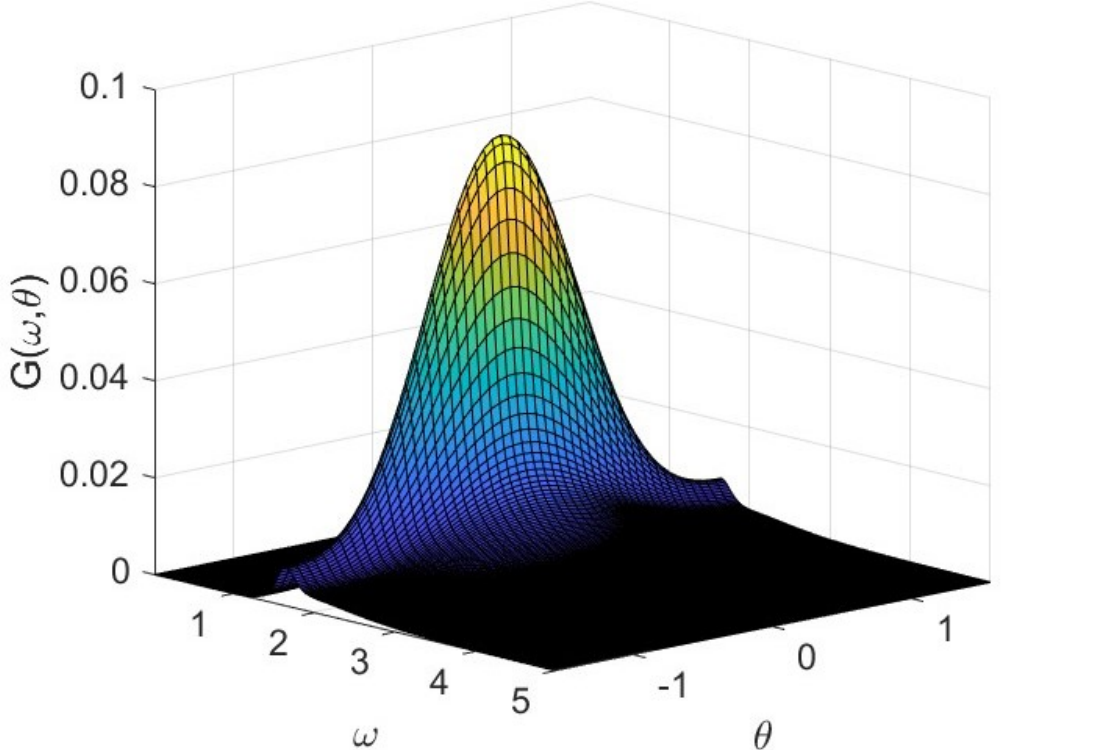}
		\end{minipage}
	}%
	
	\caption{Illustrations of 2-D and 3-D JONSWAP Spectrum under the environment with $12$ m/s wind speed and $2\times10^4$ m fetch.}
	\label{fig:JONSWAP_Spec}
	\vspace{-6mm}
\end{figure*}

% 海浪谱
The wave spectrum theory, which is commonly employed in oceanography, can describe the ocean wave by energy distribution in the frequency domain. There are various types of wave spectrum, classified according to statistics of sea conditions in specific areas.
In this paper, the JONSWAP spectrum model, proposed by Joint North Sea Wave Project, is used to introduce the ocean wave conditions \cite{A_Study_Spectra}. The 2-dimensional (2-D) JONSWAP spectrum can be formulated as
\begin{equation}\label{func1}
    S(\omega) = \frac{ag^2}{\omega^5}\exp[-1.25(\frac{\omega_p}{\omega})^4]·\gamma^{\exp[-\frac{(\omega-\omega_p)^2}{2(\sigma\omega_p)^2}]}
\end{equation}
where $\omega$ represents the frequency, and $\alpha = 0.076(\frac{U_{10}^2}{gx_f})^{0.22}$. Here $x_f$ is the fetch on the sea, $g$ represents gravity, $U_{10}$ stands for the wind speed at 10 m altitude. Moreover, $\omega_p = 22\left(\frac{g^2}{x_fU_{10}}\right)^\frac{1}{3}$ denotes the peak power, and $\gamma$ and $\sigma$ represent the shape-forming parameters.
Under 3-dimensional (3-D) circumstances, the directional spectrum is introduced to describe the ocean wave, which can be formulated as
\begin{equation}\label{func2}
    G(\omega,\theta) = \frac{1}{\pi}[1+p\cos(2\theta)+q\cos(4\theta)], \theta\le\frac{\pi}{2}
\end{equation}
where $\theta$ denotes the direction of wave propagation, and we have $p = 0.5+0.82\exp\left[-\frac{1}{2}\left(\frac{\omega}{\omega_p}\right)^4\right]
$ and $q = 0.32\exp\left[-\frac{1}{2}\left(\frac{\omega}{\omega_p}\right)^4\right]$. 
Then the 3-D JONSWAP spectrum can be calculated as 
\begin{equation}\label{func3}
    S_{\text{JONSWAP}}(\omega,\theta) = S(\omega)G(\omega,\theta)
\end{equation}
The illustrations of the 2-D and 3-D spectrum under $12$ m/s wind speed and $2\times10^4$ m fetch are exemplified as Fig. \ref{fig:JONSWAP_Spec}.

\begin{figure}[!t]
    \centering
    \includegraphics[width=1\linewidth, keepaspectratio]{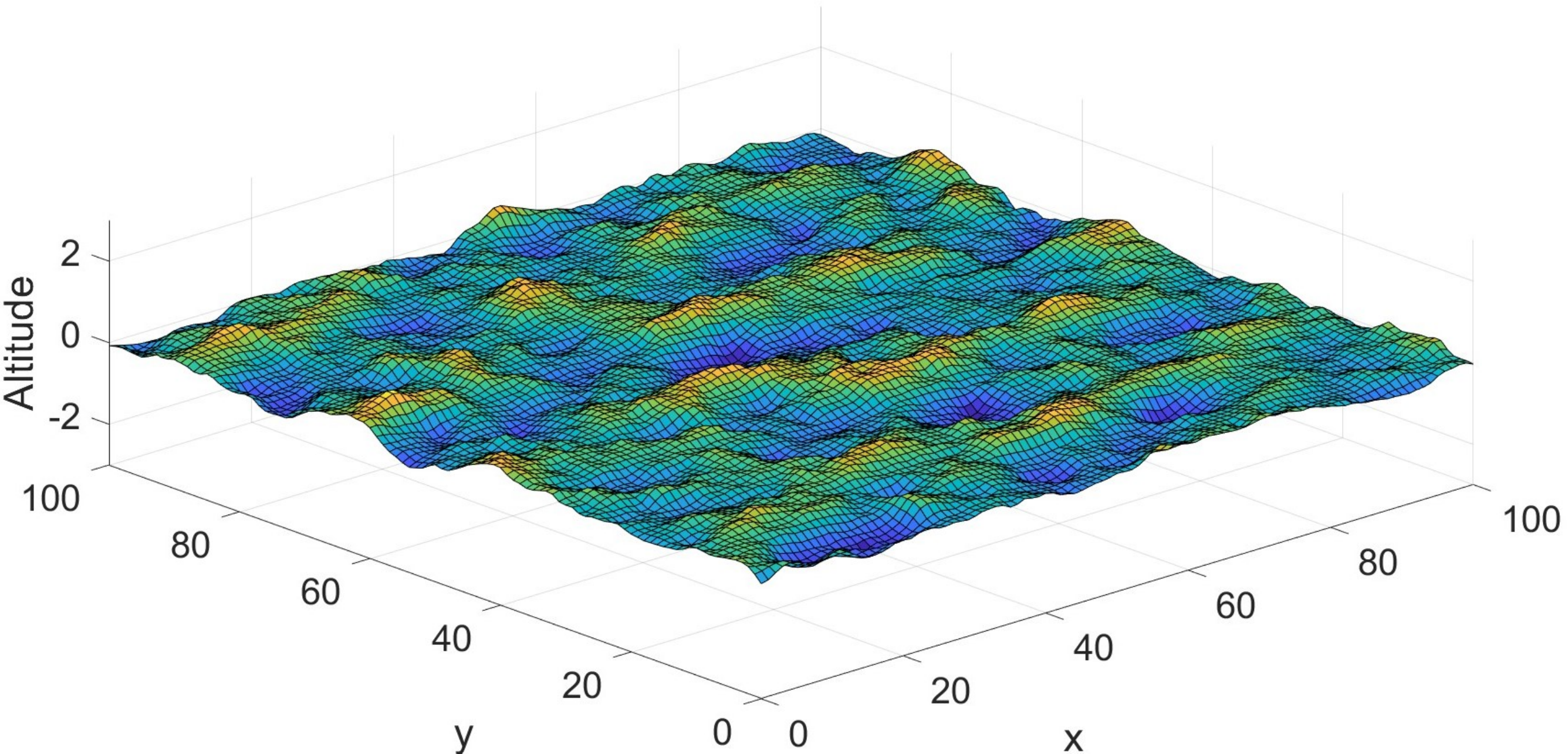}
    \caption{Simulation result of the wave surface using harmonic wave method, according to the wave spectrum in Fig. \ref{fig:JONSWAP_Spec}.}
    \label{fig:Wave_sim}
\end{figure}

% 谐波法
Based on the wave spectrum model, the harmonic wave method is employed to simulate the ocean surface with low computational cost \cite{Three-dimensional}. This method assumes that the ocean wave is composed by a group of sine functions expressed as 
\begin{equation}\label{equ:Harmonic}
T = \sum_i a_i \cos(\omega_i+\phi_i)
\end{equation}
which obeys the power distribution described by the wave spectrum. By substituting the JONSWAP spectrum into function \ref{equ:Harmonic}, the expression of the wave surface at $(x,y)$ on the plane at time instant $t$ can be formulated as
\begin{multline}\label{func4}
    T(x,y,t)=\sum_i \sum_j \sqrt{S_{\text{JONSWAP}}(\omega_i,\theta_j) d\omega d\theta} \\ \times\cos\left[ \omega_i t - \frac{\omega_i^2}{g}(x\cos\theta_j+y\sin\theta_j)+\epsilon_{i,j} \right]
\end{multline}
where $\omega_i$ and $\theta_j$ represent frequency and direction angle, respectively, and $\epsilon_{i,j}$ denotes random phase shift. The simulated model of water-air interface is exemplified as Fig. \ref{fig:Wave_sim}.

On the basis of the water-air interface model, reconstructed ray-tracing algorithm \cite{Ray_tracing} is introduced to calculate the optical propagation path 
% Specifically, we trace from the termination of each optical path to its origination, recording all of the reflection or refraction influence for channel gain calculation. 
%{\color{red}During the tracing process, the original ray tracing algorithm has been modified according to the requirements of solving the unique optical path:} 
%The original ray-tracing algorithm has been adjusted to solve the optical path 
by assuming the optical emitter has a certain extent and calculating its center by iterations. Specific steps are as follows:
The initial step is to assume a screen at the same size as the field of view (FOV) of the receiver and divide it into $m\times m$ pixels. Each pixel is represented by $(x_i,y_j,z_c)$ under receiver coordinate system, for $i,j=1,2,\ldots,m$, and $z_c$ remains a constant value. Under this condition, the coordinate difference between pixels are $\Delta x=x_i-x_{i-1}=2\frac{z_c\tan(FOV/2)}{m}$, and $\Delta y$ is the same. Meanwhile, the direction of the receiver $(0,0,z_c)$ is regarded as the central coordinate $(x_c,y_c,z_c)$.
%{\color{red}}described by $(x,y)$ coordinate.} 
Then, the ray-tracing algorithm is used to trace and calculate the light intensity of each pixel, denoted as $I_{i,j}$. 
According to the light intensity, the central coordinate is updated to the centroid of $I_{i,j}$, which can be calculated as 
\begin{equation}
 x_{c} = \frac{\sum_{i=1}^m\sum_{j=1}^m I_{i,j}x_i}{\sum_{i=1}^m\sum_{j=1}^m x_i}   
\end{equation}
and
\begin{equation}
 y_{c} = \frac{\sum_{i=1}^m\sum_{j=1}^m I_{i,j}y_j}{\sum_{i=1}^m\sum_{j=1}^m y_j}   
\end{equation}
%{\color{red}Then using the previous coordinate as the center, dividing the surrounding area into smaller units, and repeat the ray-tracing process above until the ideal accuracy is achieved. }
After updating of the central coordinate, more delicate pixel division around center $(x_c,y_c,z_c)$ is conducted for the division gap $\Delta x$ and $\Delta y$ reduce by 10 times. 
Besides, the same operation as above is repeated until the central coordinate no longer changes, which represents the receiver direction is confirmed as $\Vec{v_c}=(x_c,y_c,z_c)^T$.
Finally, since the above result $\Vec{v_c}$ is calculated in the local coordinate system of the receiver, a coordinate transformation $\Vec{v}=(x,y,z)^T=R_zR_xR_z(x_c,y_c,z_c)^T$ is performed to obtain the direction of the optical path in the absolute coordinate system \cite{lang1987linear}, thus the refraction spot and the optical path can be solved.
%\cite{x}.

% 这个海浪还有必要放吗？

\section{Beam Alignment Algorithm}\label{s4}

% 第三部分 阐述波束对准算法
% 1 基于深度强化学习的算法原理与结构
% 2 深度强化学习的模型训练
% 3 结果与评判标准
In a water-air OWC system, beam alignment between the transmitter and receiver can be mandatory to enhance the channel gain of the highly directional laser link. 
However, the unpredictability and complexity of ocean waves make it difficult for traditional model-based algorithms to handle this issue. 
Inspired by its superiority in decision-making, DRL has shown to be a promising method in beam alignment.
%Nevertheless, the development of DRL algorithms has inspired us that training DRL model is a solution to beam alignment problems.
In this paper, DDPG is selected to accomplish the beam alignment task.

\subsection{Preliminaries for DRL Techniques}\label{s4.1}
\begin{figure}[!t]
    \centering
    \includegraphics[width=0.7\linewidth, keepaspectratio]{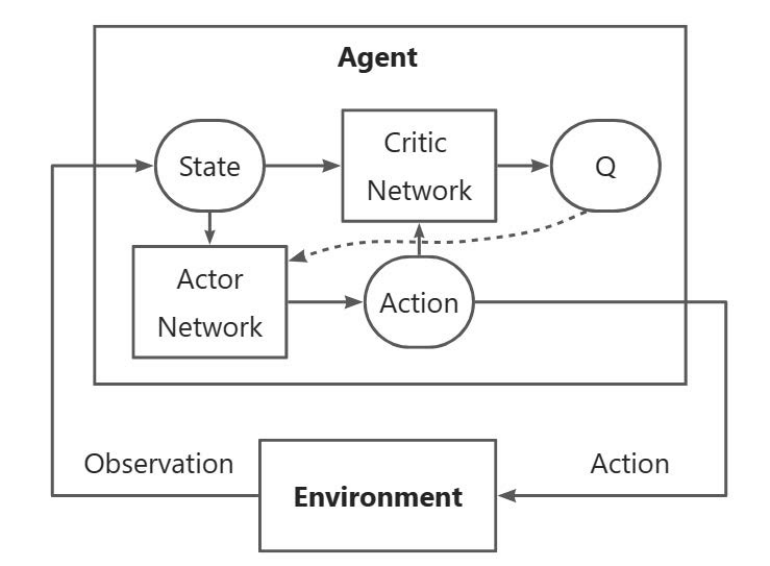}
    \caption{DDPG system model.}
    \label{fig:DDPG_model}
\end{figure}

% 简述背景与使用强化学习的原因
% In a communication environment that involves both water and air, it can be challenging for both the transmitter and receiver to accurately obtain variations of the sea surface.
In a communication environment involving both water and air, accurately obtaining variations of the sea surface can be challenging for both the transmitter and receiver.
However, reinforcement learning (RL) is a suitable algorithm that can automatically extract information from the environment. 
The main concept behind RL is to consider all of the influential factors as the $\bold{Environment}$, while an $\bold{Agent}$ is trained to observe changes in the environment parameters and conduct actions under the influence of the environment. 
% 在本文中Environment和Agent的指代
Under the proposed model, the water-air optical channel is considered as the environment, while the transmitter/receiver is regarded as the agent. 
The agent is influenced by the environment, which manifests as observation results like arrival angle and light intensity. The agent conducts actions by adjusting the transmitter and receiver direction based on the observation from the environment.
% 简述选用DDPG的特点
Furthermore, among the various algorithms in the RL area, an algorithm that can process complex changes in the environment and take continuous action is required for the problem of beam alignment under a water-air OWC channel. 
Accordingly, DDPG, a kind of acter-critic algorithm in DRL, is the most suitable algorithm for solving such problems.
% 简述DDPG的特点和性质
As shown in Fig. \ref{fig:DDPG_model}, the agent is composed of an $\bold{Actor}$ network which takes actions according to the environment, and a $\bold{Critic}$ network which judges the quality of each action and conducts feedback to regulate the actor.

\subsection{DRL Beam Alignment System Training Process}\label{s4.2}
% 训练的三要素，和与之对应的环境
%Training of DRL requires three elements, which are the environment, the agent, and training options. 
In this paper, the beam alignment algorithm is trained as the $\bold{agent}$, and the optical channel is considered as the $\bold{environment}$. Key factors that influence the training process include initial states and hyper-parameters that must also be set appropriately. 
The details of the training process will be described in the following text.

\subsubsection{Establishing of the Environment}\label{s4.2.1}
% 环境的要素
An RL Environment can be described by two functions, namely the $\bold{reset\ function}$ and the $\bold{step\ function}$. 
The reset function is operated to reset the state of the environment, which requires no extra input parameter, and generates the output of  $\bold{initial\ observation\ vector}$ and $\bold{initial\ state}$.
The step function is operated to control the changes in the environment caused by the natural deformation and the actions from the agent, which requires the $\bold{previous\ state}$ and $\bold{action\ vector}$ as input parameters, and the four output parameters of which are shown as table. \ref{table:Output_params}.
\begin{table}[!b]
    \centering
    \renewcommand{\arraystretch}{1.1}
    \caption{Output Parameters of the Step Function.}
    \resizebox{1\linewidth}{!}{
        \begin{tabular}{|l|l|}
        \hline
        Parameters       & Attribute                                      \\ \hline
        Next observation & \begin{tabular}[c]{@{}l@{}}The observation generated by the transition, caused \\ by Action, from the current state to the next one.\end{tabular} \\ \hline
        Reward           & \begin{tabular}[c]{@{}l@{}}Reward generated by the transition, caused by Action, \\ from the current state to the next one.\end{tabular}          \\ \hline
        Isdone           & \begin{tabular}[c]{@{}l@{}}Logical value indicating whether to end the \\ simulation or training episode.\end{tabular}                            \\ \hline
        Next state       & The environment state.                          \\ \hline
        \end{tabular}
    }
    \label{table:Output_params}
    \vspace{-3mm}
\end{table}
Moreover, the reset function and step function are compiled based on the cross-domain OWC channel model. 
In this research, the $\bold{reward}$ plays a crucial role in the training process of a model, influencing how quickly it converges. To achieve the optimal training result, it is necessary to design a reward function to calculate step reward based on the characteristics of the environment.
The agent tends to get more rewards by adjusting its action strategies. As the performance of beam alignment is directly related to light intensity at the receiver, the reward function should be more distinguishable by different actions. 
Therefore, the reward function is designed as a logarithm-exponential (LE) function of the light intensity. This function includes logarithm terms to distinguish intensity under small orders of magnitude and an exponential term to amplify intensity with higher value, given by
\begin{equation}\label{func}
    Reward(I) = G \cdot (ln(aI+1) + lg(bI) + \exp(cI)) + B
\end{equation}
where $I$ represents the received light intensity, $G$ is the total gain to control the peak value, $a$, $b$, and $c$ are coefficients that adjust the value of each section to the same order of magnitude, and $B$ is the bias which can sometimes simplify calculation without influencing the training process.

\subsubsection{Options of the Agent}\label{s4.2.2}
% Agent 的设置
The agent is a part of the environment and its interaction with the environment. Therefore, the interface between the agent and the environment is a key impact factor in both the training process and practical application. 
% obs 和 act
The interface between the agent and the environment can be divided into two groups, which are organized into two vectors named observation and action. 
% observation
The observation vector contains all factors that can be observed by the agent from the environment and acts as the input of the action and critic network. 
In a cross-domain OWC channel, the observation vector is designed as an eight-dimensional vector including the transmitter direction, the receiver direction, the light intensity, and relative time.
% action
The action vector contains all factors that decide the action of the agent. 
In a cross-domain OWC channel, the observer vector is designed as a 4-dimensional vector including the transmitter direction and the receiver direction. 
It is worth noting that the orientation of the transmitter and receiver is limited (transmitter upwards and receiver downwards), so two-dimensional vectors can be used to represent the three-dimensional direction they can reach. Such operation can reduce network parameters, thereby reducing training and running costs.

\begin{table}[!b]
    \centering
    \caption{Hyper-Parameters for DRL Training.}
    \resizebox{0.75\linewidth}{!}{
    \begin{tabular}{|c|c|}
    \hline
        \textbf{Hyper-parameters} & \textbf{Values} \\ \hline
        Noise & Gaussian Action Noise \\ \hline
        Discount Factor & 0.2 \\ \hline
        Sample Time & 0.05 s \\ \hline
        Buffer Length & 1e6 \\ \hline
        Mini Batch Size & 64 \\ \hline
        Actor Learning Rate & 1e-3 \\ \hline
        Critic Learning Rate & 1e-4 \\ \hline
        Max Episode & 500 \\ \hline
        Max Steps per Episode & 500 \\ \hline
    \end{tabular}
    }
    \label{table:Training_params}
    \vspace{-5mm}
\end{table}

\subsubsection{Hyper-Parameters for Training Process}\label{s4.2.3}
% 描述训练参数的影响
After the environment and agent have been established and set, the training process is ready to begin. Before training, various hyper-parameters can be set to influence the training process and result from various aspects. 
Part of the hyper-parameters that have a critical influence on the training process are customized and shown in the table. \ref{table:Training_params}, while others remain default value.

\section{Simulation Results}\label{s5}
In this section, the results of the simulation are displayed and analyzed. 
% 第四部分 阐述实验的结果
% 1 实验参数设定与结果呈现
Parameter settings of the simulation environment are as follows, the sea level is regarded as $0m$ and the transmitter and receiver are assumed to be located at $-10m$ in the sea and $10m$ in the air. The refresh interval of the environment is set to $0.05s$. 
During the experiment, the proposed method is compared to these methods below, $a)$ the theoretical upper-bound (UB) with the maximum channel gain, $b)$ straight-facing alignment strategy that the transmitter and receiver face directly to each other, $c)$ gain with no alignment algorithm.

% 2 与其他算法对比
\begin{figure}[!t]
    \centering
    \includegraphics[width=0.9\linewidth, keepaspectratio]{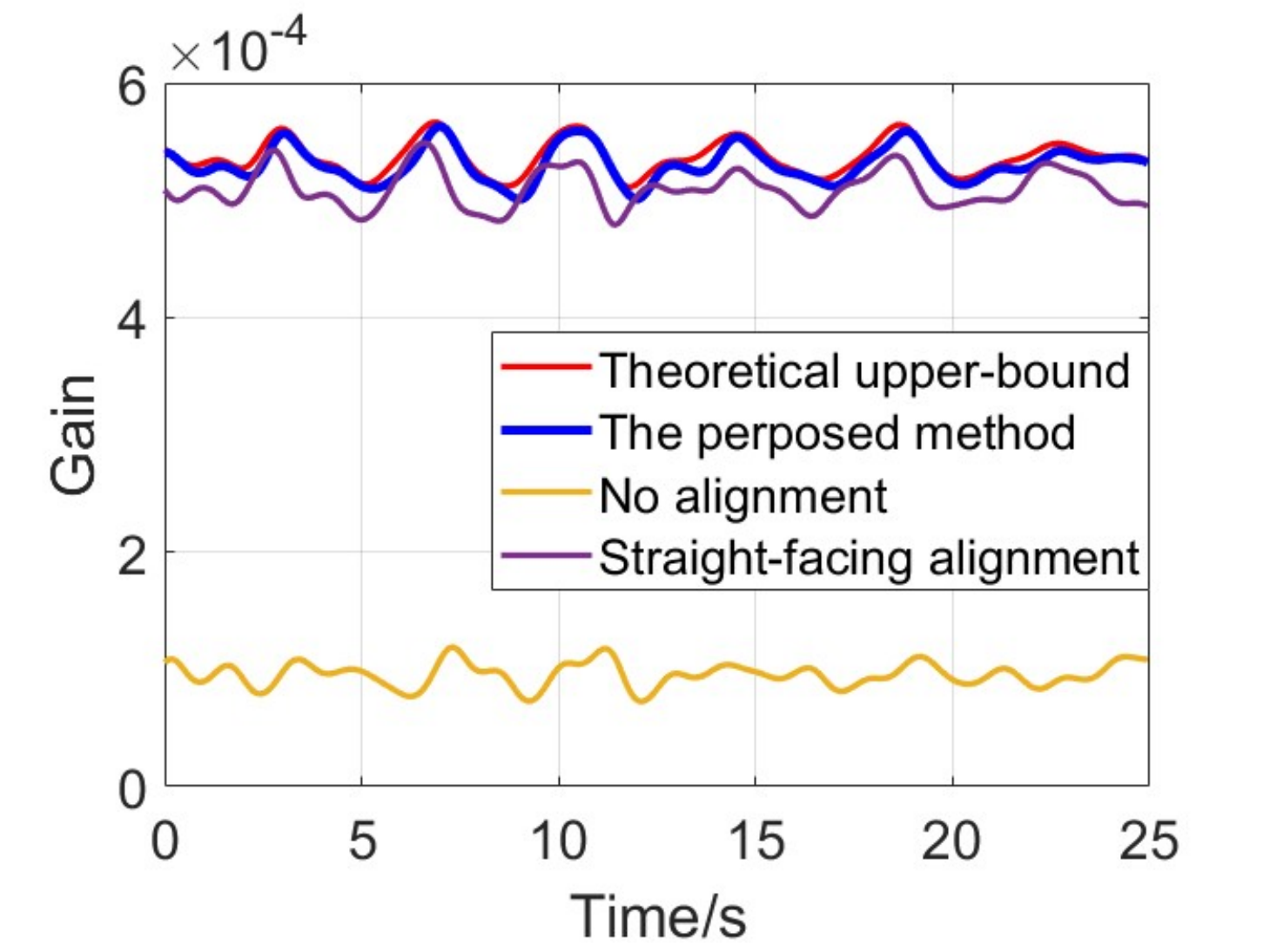}
    \caption{Comparison of OWC channel gain over time between the proposed DRL-based beam alignment scheme and other counterparts.}
    \label{fig:Channel_Gain}
    \vspace{-3mm}
\end{figure}

% 按时间仿真，说明了方法的优越性和稳定性
The operation simulation of the proposed method is conducted under an environment that includes a wind speed of $12m/s$, $20m$ vertical distance, $10\%$ horizontal offset, and a time-span of $25s$ with $0.05s$ sampling interval.
The simulation result describes the OWC channel gain over time as Fig. \ref{fig:Channel_Gain}. 
The proposed method performs much better than the method without alignment and is closer to the theoretical upper-band than the straight-facing alignment method.

\begin{figure}[!t]
	\centering
	\subfigure[Average of channel gain.]{
		\begin{minipage}[t]{1\linewidth}
			\centering
			\includegraphics[width=0.85\linewidth]{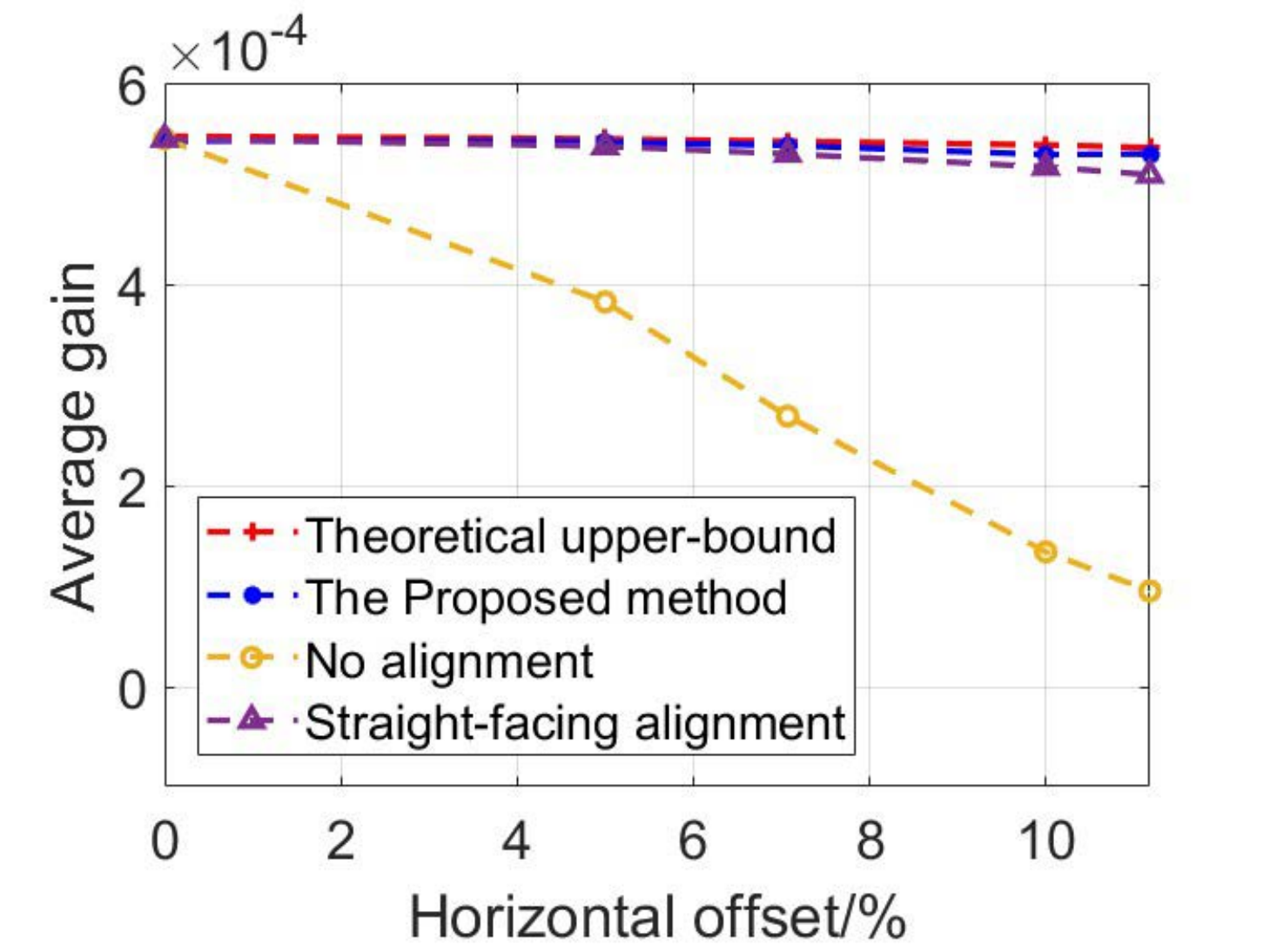}
		\end{minipage}
	}%
 
	\subfigure[$\bold{\sigma_{diff}^2}$ of different methods.]{
		\begin{minipage}[t]{1\linewidth}
			\centering
			\includegraphics[width=0.85\linewidth]{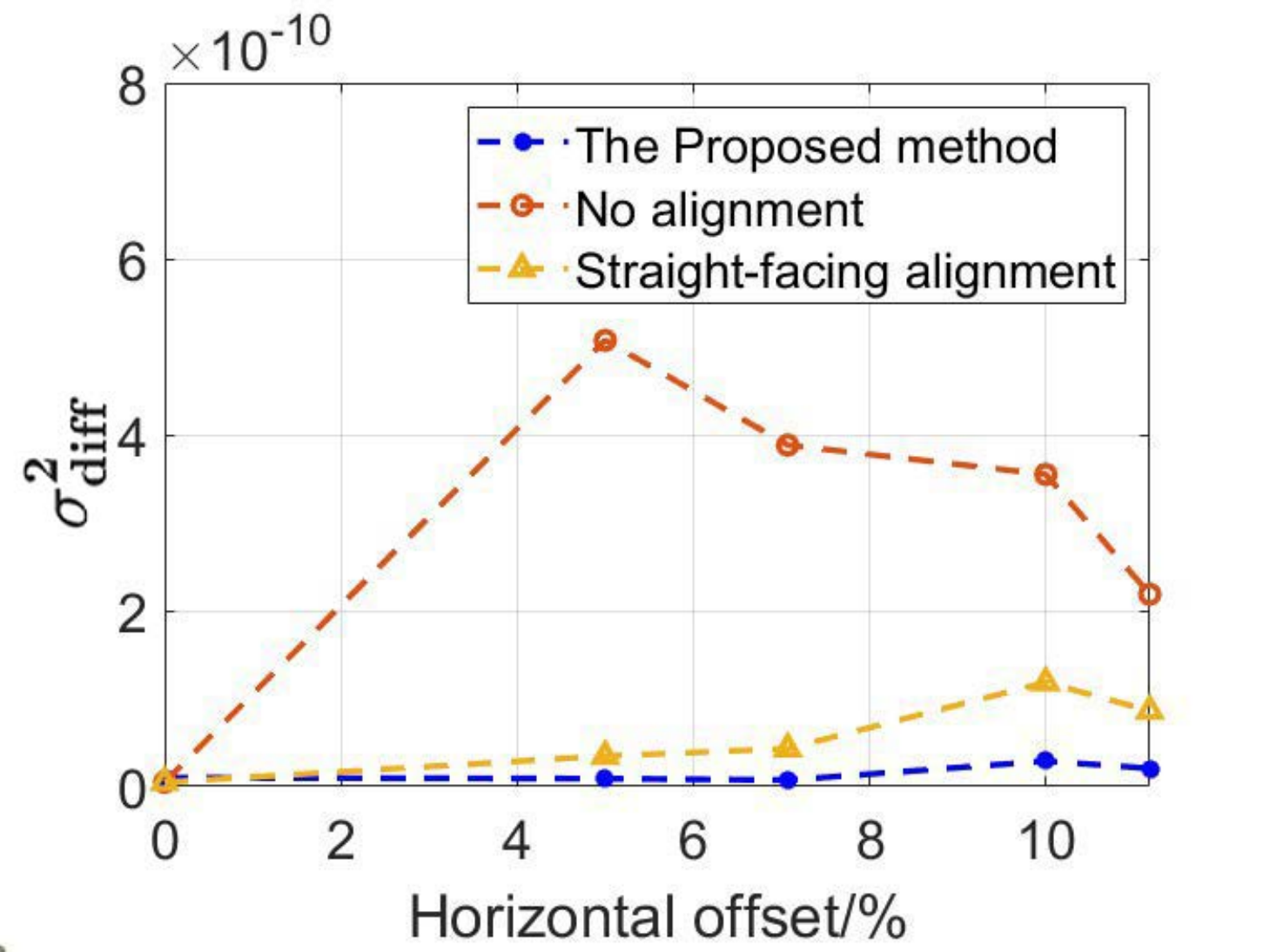}
		\end{minipage}
	}%
	
	\caption{The average channel gain and $\bold{\sigma_{diff}^2}$ that are achieved by the proposed DRL-based beam alignment scheme, in comparison with different counterparts.}
 %Simulation result of OWC channel gain average value and variance over horizontal offset, the proposed method compares with other different methods.
	\label{fig:Offset}
	\vspace{-5mm}
\end{figure}

% 水平偏差仿真，抗水平偏差的能力
%The proposed method performs better at dealing with horizontal offset. 
%Such capability is verified in subsequent simulations, where beam alignment algorithms are operated under different horizontal offsets.
As shown in Fig. \ref{fig:Offset}(a), the average channel gain of the proposed algorithm maintains a high level which is closer to the theoretical upper-bound than other methods, and takes little influence from the horizontal offset. 
In addition, to verify the stability of the method, a variable $\bold{\sigma_{diff}^2=var}(G_{UB}-G)$ is used to measure the stability, where $G_{UB}$ and $G$ are the channel gain of the upper-bound and each method, respectively. A lower value of $\bold{\sigma_{diff}^2}$ indicates that the method is more stable against wave influences. As shown in Fig. \ref{fig:Offset}(b), $\bold{\sigma_{diff}^2}$ of the proposed method is lower than its counterparts, which indicates the higher stability of the proposed method.
Overall, the proposed method has better performance in terms of resistance to influences of horizontal offset, as well as influences caused by dynamic-wave fluctuations, thus having higher channel gain and better stability in the water-air OWC channel when compared to other methods.

% 
%The article also researched the channel gain over different wind speed which affected the fluctuation level of the ocean waves. The simulation shows that the mean value and variance of the channel gain over the wind speed at $10m$ above the sea level from $4m/s$ to $20m/s$. As shown in Fig. \ref{}, the 

% 3 结果分析

\section{Conclusion}\label{s6}\vspace{-1mm}
In this paper, we propose a DRL-based beam alignment strategy for water-air direct OWC, which can significantly enhance the resilience to dynamic characteristics of the water surface. More specifically, the dynamic properties of the water-air interface are investigated with sea-wave spectrum analysis, followed by the propagation modeling using ray-tracing methods. On the basis of the established channel model, the beam alignment problem is modeled as a reinforcement learning process, where the DDPG scheme is employed. To further enhance its convergence performance, a logarithm-exponential (LE) nonlinear reward function with respect to the received signal strength is developed for more distinguishable rewards between different actions. Simulation results demonstrate that %xxx
the proposed method keeps high channel gain in the water-air OWC channel and can resist influences of wave fluctuations and horizontal offsets.

% To address the issue, this article proposes one novel beam alignment strategy based on deep reinforcement learning (DRL) for water-air direct optical wireless communications. Specifically, the dynamic water-air interface is mathematically modelled using sea-wave spectrum analysis, followed by characterization of the propagation channel with ray-tracing techniques. Then the deep deterministic policy gradient (DDPG) scheme is introduced for DRL-based transceiving beam alignment. A logarithm-exponential (LE) nonlinear reward function with respect to the received signal strength is designed for high-resolution rewarding between different actions. Simulation results validate the superiority of the proposed DRL-based beam alignment scheme.

% 全文总结

\vspace{-1mm}
\section*{Acknowledgment}\vspace{-1mm}
This work was supported in part by the National Natural Science Foundation of China under Grant No. 62088101, in part by the Young Elite Scientists Sponsorship Program by CAST under Grant 2022QNRC001, in part by the National Natural Science Foundation of China under Grant 62101306, and in part by the National Natural Science Foundation of China under Grant 62371065. \emph{(Jiayue Liu and Tianqi Mao are Co-first authors with equal contribution.) (Corresponding author: Dezhi Zheng.)}

%\bibliographystyle{ieeetr}
%\bibliography{Reference}

\begin{thebibliography}{10}
\providecommand{\url}[1]{#1}
\csname url@samestyle\endcsname
\providecommand{\newblock}{\relax}
\providecommand{\bibinfo}[2]{#2}
\providecommand{\BIBentrySTDinterwordspacing}{\spaceskip=0pt\relax}
\providecommand{\BIBentryALTinterwordstretchfactor}{4}
\providecommand{\BIBentryALTinterwordspacing}{\spaceskip=\fontdimen2\font plus
\BIBentryALTinterwordstretchfactor\fontdimen3\font minus \fontdimen4\font\relax}
\providecommand{\BIBforeignlanguage}[2]{{%
\expandafter\ifx\csname l@#1\endcsname\relax
\typeout{** WARNING: IEEEtran.bst: No hyphenation pattern has been}%
\typeout{** loaded for the language `#1'. Using the pattern for}%
\typeout{** the default language instead.}%
\else
\language=\csname l@#1\endcsname
\fi
#2}}
\providecommand{\BIBdecl}{\relax}
\BIBdecl

% 1
\bibitem{An_overview}
\BIBentryALTinterwordspacing
M.~C. Domingo, ``An overview of the internet of underwater things,'' \emph{Journal of Network and Computer Applications}, vol.~35, no.~6, pp. 1879--1890, 2012. [Online]. Available: \url{https://www.sciencedirect.com/science/article/pii/S1084804512001646}
\BIBentrySTDinterwordspacing

% 2
\bibitem{Overview_UWC}
\BIBentryALTinterwordspacing
------, ``Overview of channel models for underwater wireless communication networks,'' \emph{Physical Communication}, vol.~1, no.~3, pp. 163--182, 2008. [Online]. Available: \url{https://www.sciencedirect.com/science/article/pii/S1874490708000451}
\BIBentrySTDinterwordspacing

% 3
\bibitem{Zhou_access_19}
J.~Zhou, H.~Jiang, P.~Wu, and Q.~Chen, ``Study of propagation channel characteristics for underwater acoustic communication environments,'' \emph{IEEE Access}, vol.~7, pp. 79\,438--79\,445, 2019.

% 4
\bibitem{A_Survay_UOWC}
Z.~Zeng, S.~Fu, H.~Zhang, Y.~Dong, and J.~Cheng, ``A survey of underwater optical wireless communications,'' \emph{IEEE Communications Surveys \& Tutorials}, vol.~19, no.~1, pp. 204--238, 2017.

% 5
\bibitem{Waving_Effect}
T.~Lin, C.~Fu, T.~Wei, N.~Huang, X.~Liu, L.~Tang, L.~Su, J.~Luo, and C.~Gong, ``Waving effect characterization for water-to-air optical wireless communication,'' \emph{Journal of Lightwave Technology}, vol.~41, no.~1, pp. 120--136, 2023.

% 6
\bibitem{Recent_Progress}
H.~Luo, J.~Wang, F.~Bu, R.~Ruby, K.~Wu, and Z.~Guo, ``Recent progress of air/water cross-boundary communications for underwater sensor networks: A review,'' \emph{IEEE Sensors Journal}, vol.~22, no.~9, pp. 8360--8382, 2022.

% 7
\bibitem{xiaoyang2019performance}
X.~Xiaoyang, S.~Liwei, Z.~Jinyu, Z.~Wu, D.~Wenjing, Z.~Xu, ``Performance analysis of sea unmanned ship routing protocol based on ad hoc network,'' in \emph{2019 International Conference on Information Technology and Computer Application (ITCA)}.\hskip 1em plus 0.5em minus 0.4em\relax IEEE, 2019, pp. 221--224.

% 8
\bibitem{Underwater_optic}
H.~Kaushal and G.~Kaddoum, ``Underwater optical wireless communication,'' \emph{IEEE Access}, vol.~4, pp. 1518--1547, 2016.

% 9
\bibitem{Effect_of}
Y.~Dong, S.~Tang, X.~Zhang, ``Effect of random sea surface on downlink underwater wireless optical communications,'' \emph{IEEE Communications Letters}, vol.~17, no.~11, pp. 2164--2167, 2013.

% 10
\bibitem{Underwater_and}
L.-K. Chen, Y.~Shao, and Y.~Di, ``Underwater and water-air optical wireless communication,'' \emph{Journal of Lightwave Technology}, vol.~40, no.~5, pp. 1440--1452, 2022.

% 11
\bibitem{Mitigation}
Y.~Di, Y.~Shao, and L.-K. Chen, ``Mitigation of wave-induced packet loss for water-air optical wireless communication by a tracking system,'' in \emph{2021 Optical Fiber Communications Conference and Exhibition (OFC)}, 2021, pp. 1--3.

% 12
\bibitem{Preliminary}
T.~Lin, N.~Huang, C.~Gong, J.~Luo, Z.~Xu, ``Preliminary characterization of coverage for water-to-air visible light communication through wavy water surface,'' \emph{IEEE Photonics Journal}, vol.~13, no.~1, pp. 1--13, 2021.

% 13
\bibitem{Improvement_of}
D.~Wu, Z.~Ghassemlooy, H.~L. Minh, S.~Rajbhandari, and A.~C. Boucouvalas, ``Improvement of the transmission bandwidth for indoor optical wireless communication systems using a diffused gaussian beam,'' \emph{IEEE Communications Letters}, vol.~16, no.~8, pp. 1316--1319, 2012.

% 14
\bibitem{Study_on}
H.~Wu and Q.~Fan, ``Study on led visible light communication channel model based on poisson stochastic network theory,'' in \emph{2020 International Conference on Wireless Communications and Smart Grid (ICWCSG)}.\hskip 1em plus 0.5em minus 0.4em\relax IEEE, 2020, pp. 5--9.

% 15
\bibitem{Absorption_spec}
R.~M. Pope and E.~S. Fry, ``Absorption spectrum (380--700 nm) of pure water. ii. integrating cavity measurements,'' \emph{Applied optics}, vol.~36, no.~33, pp. 8710--8723, 1997.

% 16
\bibitem{Fresnel}
A.~I. Lvovsky, ``Fresnel equations,'' \emph{Encyclopedia of Optical Engineering}, vol.~27, pp. 1--6, 2013.

% 17
\bibitem{A_Study_Spectra}
J.~Prendergast, M.~Li, and W.~Sheng, ``A study on the effects of wave spectra on wave energy conversions,'' \emph{IEEE Journal of Oceanic Engineering}, vol.~45, no.~1, pp. 271--283, 2020.

% 18
\bibitem{Three-dimensional}
Z.~Chang, F.~Han, Z.~Sun, Z.~Gao, and L.~Wang, ``Three-dimensional dynamic sea surface modeling based on ocean wave spectrum,'' \emph{Acta Oceanologica Sinica: English ver.}, vol.~40, no.~10, p.~11, 2021.

% 19
\bibitem{Ray_tracing}
C.~Benthin, I.~Wald, M.~Scherbaum, and H.~Friedrich, ``Ray tracing on the cell processor,'' in \emph{2006 IEEE Symposium on Interactive Ray Tracing}, 2006, pp. 15--23.

% 20
\bibitem{lang1987linear}
S.~Lang, \emph{Linear algebra}.\hskip 1em plus 0.5em minus 0.4em\relax Springer Science \& Business Media, 1987.


\end{thebibliography}
\vspace{-1mm}

%%

\clearpage

\end{document}